# Tuning the thermoelectric properties of SrTiO$_3$ by controlled oxygen doping.

P. L. Bach,[1] V. Leborán,[1] V. Pardo,[2,3] A. S. Botana,[2,3] D. Baldomir,[2,3] F. Rivadulla[1*]

[1]*Centro de Investigación en Química Biológica y Materiales Moleculares (CIQUS), Universidade de Santiago de Compostela, E-15782, Santiago de Compostela, Spain.*
[2]*Departamento de Física Aplicada, Universidade de Santiago de Compostela, E-15782, Santiago de Compostela, Spain.*
[3]*Instituto de Investigacións Tecnolóxicas, Universidade de Santiago de Compostela, E-15782, Santiago de Compostela, Spain.*

**Abstract**

We report the thermoelectric properties (Seebeck coefficient, thermal conductivity, and electrical resistivity) of lightly doped single crystals of (001)-oriented SrTiO$_3$ (STO). Hall effect measurements show that electron doping around $10^{-5}$ carriers per unit cell can be achieved by vacuum annealing of the crystals under carefully controlled conditions. The steep density of states near the Fermi energy of STO at this doping level (confirmed by ab initio calculations) retains an unusually large Seebeck coefficient, in spite of an increase in the electronic conductivity by several orders of magnitude. This effect, combined with a decrease in thermal conductivity due to vacancy disorder scattering makes intrinsic doping in STO (and other materials) an alternative strategy to optimize its thermoelectric figure of merit.

Thermoelectric (TE) materials with a large figure of merit, $Z=S^2/\rho\kappa$ (S, $\rho$, and $\kappa$ are the Seebeck coefficient, the electrical resistivity and the thermal conductivity, respectively), have attracted much interest because of their potential practical applications.[1] The most serious difficulty in the enhancement of the TE performance to a desirable value ZT=2-4 is the interdependence of S, $\rho$ and $\kappa$; the optimization of one of them normally results in the other evolving in the undesired direction.[2]

Classic strategies to pass this barrier and improve ZT rely on the use of alloys and heavy-element-based degenerate semiconductors, in which the thermal conductivity is severely reduced.[3] Also, the fabrication of intrinsically inhomogeneous systems, like AgPb$_{18}$SbTe$_{20}$ (ZT~2.4, T≈800 K),[4,5] opens a path for optimizing the TE performance through the search for a controlled method to fabricate "inhomogeneous" materials.

In the last few years, nanostructuring of classical materials has emerged as an alternative route to improve the TE performance.[6] Although the possibility of bulk production is normally preferred due to the ease of synthesis and scalability, physical deposition techniques (molecular beam epitaxy, laser ablation, sputtering, etc.) allow the fabrication of nanostructures in which dimensionality and composition can be fine tuned to explore the effects of quantum confinement in the performance of TE devices.[6] By this approach, ZT≈2 has been reported at room temperature in supperlattices of PbTe nanodots,[7] and large enhancements of the power factor have also been found in thin-films and multilayers of both amorphous and crystalline semiconductors,[8,9,10] Si nanowires,[11] etc. Ohta *et al.*[12] reported an enhanced ZT in the two-dimensional (2D) electron gas that forms at the interface of ultrathin Nb-doped SrTiO$_3$ (STO), sandwiched between insulating STO barriers.

On the other hand, oxygen vacancies can be easily induced in STO by high temperature vacuum annealing or ion-beam etching, resulting in n-type doping.[13,14,15,16,17]

If an strategy for controlled oxygen removal could be designed, we could reach doping levels that are unattainable by traditional chemical methods.

Here we show that the number of charge carriers can be controlled by careful high temperature vacuum annealing protocol, tuning the electron density through several orders of magnitude. We explore the possibility of exploiting this mechanism of intrinsic doping to optimize the TE performance of STO.

Square, flat, (001)-oriented STO crystals (5×5×0.5 mm$^3$) were annealed in a high vacuum chamber between 760 $^o$C and 850 $^o$C for either 30 or 60 minutes at oxygen pressures ranging from $10^{-3}$ Torr to $10^{-7}$ Torr. For resistivity and Hall effect measurements, Pt/Au pads were evaporated at the corners of the crystals, in a van der Pauw configuration. Resistances, both R$_{xx}$ and R$_{xy}$, were measured by taking I-V curves at each temperature using a Keithley 4200 SCS. Thermal measurements were done in a steady-state configuration. Ab initio calculations were performed on cubic perovskite STO based on a density functional theory framework[18] using the WIEN2k package.[19] Calculations were fully converged with respect to all the parameters considered using RK$_{max}$=7 and a k-mesh of 12 x 12 x 12. Transport properties were calculated using a semiclassical approach based on Boltzmann's equation within the constant scattering time approximation. For these calculations we used an enlarged k-mesh of 40 x 40 x 40, enough to yield fully converged results. For the exchange-correlation potential, we used the Tran-Blaha-modified Becke-Johnson (TB-mBJLDA) potential,[20] that allows a fully ab initio, parameter-free description of the system. This potential yields an improved agreement in the calculated band gap and transport properties of both semiconductors and strongly correlated electron systems.[21,22]

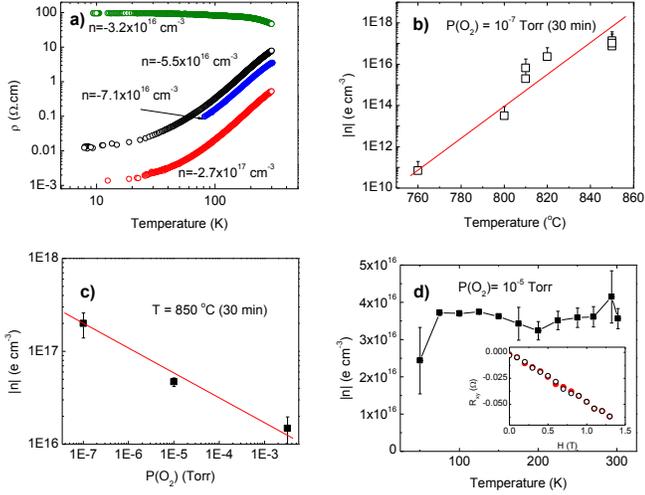

Figure 1: (a) Resistivity vs temperature for representative samples annealed at different oxygen pressures. The blue curve (n= -7.1 × 10$^{16}$ cm$^{-3}$) corresponds to a sample annealed under the same conditions as the black curve (n= -5.5 × 10$^{16}$ cm$^{-3}$), but twice the time. The numbers represent the carrier density measured from the Hall effect. The dependence of the final carrier density on annealing temperature, for a fixed oxygen pressure, is shown in (b) and in (c) for a fixed temperature and different oxygen pressures. (d) The temperature dependence of the charge carrier density determined from Hall effect, obtained by fitting the Hall resistance ($R_H$) field sweeps data (inset).

Resistivity and Hall effect measurements (Fig. 1) show that the number of charge carriers, and hence the electronic conductivity of STO, can be tuned by a combination of oxygen pressure and annealing temperature, over a range which spans approximately six orders of magnitude of charge carrier densities (Fig. 1b and 1c). We noted that these changes are reversible; the fully insulating behavior is recovered by re-annealing conductive substrates according to the same procedure, but under oxygen pressure, ≈200 mTorr. We have also observed that annealing in a non-oxygen atmosphere (N$_2$) is equivalent to vacuum annealing at a comparable oxygen partial pressure. Furthermore, increasing the annealing time of the crystals produces only a modest increase in charge carrier density (Fig. 1a), showing that annealing at 850 °C for 30 min allows the oxygen concentration in STO to nearly reach its equilibrium value with the vacuum chamber's oxygen partial pressure, suggesting that oxygen has had time to diffuse from deeper layers. Indeed, the c-axis resistance between the front and back faces of the crystal is comparable to the in-plane resistance.

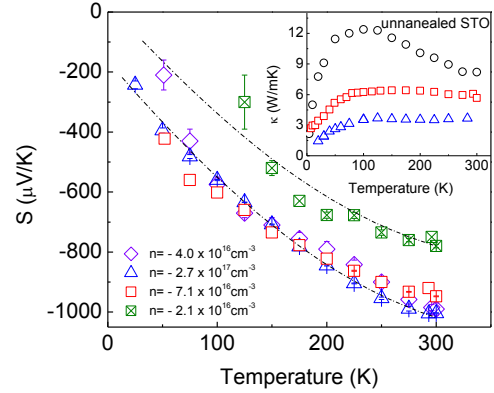

Figure 2: Temperature dependence of the thermopower and thermal conductivity (inset) of vacuum annealed STO. For the annealed samples, the thermal conductivity has been corrected for radiation losses.[23,24] Lines are a guide to the eye. The same symbols are used in the inset as for the samples in the main panel.

Thermoelectric power and thermal conductivity were also measured for crystals annealed under different conditions. The former (Seebeck coefficient) is very sensitive to the dimensionality of the system.[25] In fact, an enhanced Seebeck effect has been reported in 2D electron gases at LaAlO$_3$/SrTiO$_3$ heterointerfaces[26] and ultrathin layers of Nb-doped STO sandwiched between insulating STO barriers.[12] However, we show in Fig. 3 that STO can be precisely doped via oxygen vacancies to show a Seebeck coefficient comparable to that reported in 2D heterointerfaces. Therefore, this effect seems to be related to the low doping level and the band structure of bulk STO rather than being caused by 2D confinement.

An expression for the Seebeck coefficient of 3D metals and degenerate semiconductors can be derived from Boltzmann's equation:[27]

$$S = \frac{\pi^2 k_B^2 T}{3e} \frac{8m^*}{h^2} \left(\frac{\pi}{3n}\right)^{2/3} (1+r)$$

where $n$ is the electron density, and $r$ comes from the energy dependence of the scattering time, $\tau = \tau_0 E^r$. The band structure for cubic STO (see Fig. 3a) has a largely anisotropic mass tensor around the Γ point in the electron-doping region, with a large component along the (100) direction for one of the t$_{2g}$ bands (the manifold splits into two light and one heavy electron bands along that direction), whereas all the other values of the mass tensor are around m$^*$ = 3 m$_e$. Using a value of r = 2 (3D phonon scattering), and m$^*$ = 3 m$_e$, we obtain S≈1300 µV/K at 300 K for n=-3×10$^{17}$ e/cm$^3$. Although this is a very rough calculation (the actual proportion of light/heavy electrons is not taken into account), it shows that the band structure of STO is able to account for the unusual combination of high thermopower and large electronic mobility at small doping.

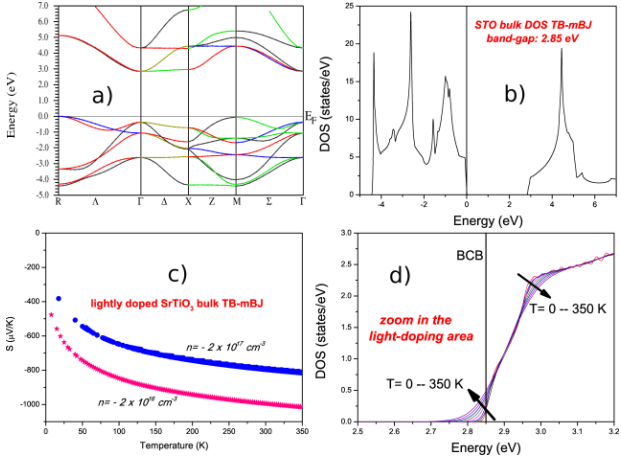

Figure 3. Summary of results from band structure and transport properties calculations. Panel a) shows the band structure of cubic $SrTiO_3$, b) density of states, c) Seebeck coefficient calculated for two doping values, d) zoomed density of states around the bottom of the conduction band at various temperatures (BCB stands for bottom of the conduction band).

In order to check the validity of this conclusion, the band structure of STO was calculated ab initio, and the density of states at the actual doping level was used to obtain the temperature dependence of the Seebeck coefficient using Boltzmann's transport theory (Fig. 3). Our calculations focused on obtaining the thermopower of bulk cubic perovskite STO at very low doping values, simulating the doping induced by oxygen vacancies, *i.e.* placing the chemical potential very close to the bottom of the conduction band. The doping values we are interested in are about one oxygen vacancy per $10^5$-$10^6$ unit cells, which are not attainable via supercells or even the virtual crystal approximation. Thus, calculations on the undoped material were used instead and the chemical potential was displaced accordingly to yield the appropriate charge carrier concentration. Calculations are summarized in Fig. 3. The TB-mBJLDA potential yields a substantially improved band gap over standard functionals based on the local density approximation. The value we obtain is 2.85 eV, which agrees reasonably well with the experimental value of about 3.25 eV.[28] In Fig. 3d we show in detail the density of states in the interesting small electron-doping region (close to the bottom of the conduction band) with various curves at different temperatures to see the broadening effects in the density of states as temperature increases. Fig. 3c shows the thermopower as a function of temperature for fixed values of the carrier concentration. Since Hall experiments show the number of carriers remains fairly constant in the 100 – 300 K temperature interval (Fig. 1d), we have fixed the carrier concentration at two values: $-2\times10^{16}$ and $-2\times10^{17}$ $cm^{-3}$, and calculated the Seebeck coefficient vs. temperature. We find large values at such very low doping, in good agreement with experiment. This is a consequence of being very close to the edge of the conduction band of a wide-gap insulator. The value of the thermopower decreases as doping increases, as one would expect, but remains high as long as this very low electron doping is retained. We do however note that our measured thermopower exhibits less suppression with doping than the theoretical calculations suggest.

This doping level is impossible to obtain by traditional chemical methods (*i.e.* by replacement of $Sr^{2+}$ with $La^{3+}$),[29] so that intrinsic electron doping by a fine tuning of oxygen vacancies opens new ways to control the transport properties of bulk oxides.

With respect to thermal transport, the results in Fig. 2 (inset) show that phonon scattering by oxygen vacancies suppresses the broad phonon peak and reduces the thermal conductivity.[30] The combination of a suppressed κ with a large Seebeck coefficient at low doping results in a substantial increase of the TE figure of merit (Fig. 4).

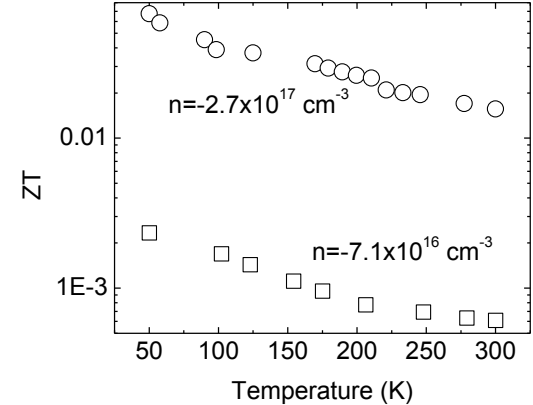

Figure 4: Temperature dependence of the dimensionless thermoelectric figure of merit for two samples.

Hence, large Seebeck values can be maintained in oxygen deficient STO, while simultaneously reducing the thermal conductivity and electrical resistivity. This, then, suggests that the TE figure of merit for STO might be increased significantly by a carefully tuned annealing/doping protocol. Indeed, the ratio of the ZT for crystals annealed at $10^{-7}$ Torr with respect to $10^{-5}$ Torr is a factor ≈20 larger at room temperature, although still far from being useful for practical applications.

In summary, the ability of STO to easily accept oxygen vacancies can be exploited to tune its electronic properties, through the modification of several decades in charge carrier density. Moreover, the small sensitivity of the thermoelectric power to the electron density at this doping level, and the reduction of the thermal conductivity due to vacancy scattering, represents a controllable strategy to improve ZT.

**Acknowledgements**

We acknowledge financial support from ERC (StG-259082, 2DTHERMS), and Ministerio de Economía y Competitividad of Spain (MAT2010-16157). V.P. and A.S.B. acknowledge the Spanish Government for support through the Ramón y Cajal and FPU program, respectively. Financial support was obtained from the Ministry of Science of Spain through Project No. MAT-200908165.